\documentclass[a4paper]{jpconf}

\newcommand{\be}{\begin{equation}}
\newcommand{\ee}{\end{equation}}
\newcommand{\bea}{\begin{eqnarray}}
\newcommand{\eea}{\end{eqnarray}}
\newcommand{\ben}{\begin{eqnarray*}}
\newcommand{\een}{\end{eqnarray*}}
\newcommand{\doa}{\downarrow}
\newcommand{\upa}{\uparrow}
\newcommand{\average}[1]{\langle #1 \rangle}
\newcommand{\sign}{\textrm{sign}}

\newcommand{\nmodel}{\textit{U-V-J}}
\newcommand{\eg}{\textit{e.g.}}

\usepackage{graphicx,color}
\usepackage{amssymb}
\usepackage{amsmath}
\usepackage{url}
\usepackage[update,prepend]{epstopdf}
\usepackage{hyperref}
\usepackage{bm}
\usepackage[T1]{fontenc}

\begin{document}

\title{Spin and charge orderings in the atomic limit
of the~\nmodel~model}

\author{F Mancini$^{1,2}$, E Plekhanov$^{1,2,3}$, and G Sica$^{1}$}

\address{$^1$ Dipartimento di Fisica ``E.R. Caianiello'', Universit\`a
degli Studi di Salerno, 84084 Fisciano (SA), Italy}

\address{$^2$ Unit\`a CNISM di Salerno, Universit\`a degli Studi di
Salerno, 84084 Fisciano (SA), Italy}

\address{$^3$ CNR-SPIN, UoS di Salerno, 84084 Fisciano (SA), Italy}

\ead{plekhanoff@physics.unisa.it}

\begin{abstract}
In this paper we study a generalization of the 1D Hubbard model by
considering density-density and Ising-type spin-spin nearest neighbor
(NN) interactions, parameterized by $V$ and $J$, respectively. We present the
$T=0$ phase diagram for both ferro ($J>0$) and anti-ferro ($J<0$)
coupling obtained in the narrow-band limit by means of an extension to
zero-temperature of the transfer-matrix method. Based on the values of
the Hamiltonian parameters, we identify a number of phases that involve
orderings of the double occupancy, NN density and spin
correlations, being these latter very fragile.
\end{abstract}

\section{Introduction}
The Hubbard model (HM) is undoubtedly one of the most studied model in
condensed matter physics. In one spatial dimension (1D), the exact Bethe
solution for the pure HM is known~\cite{bethe}. During its long history,
several extensions to the HM have been proposed. Among them, the
inclusion of non-local charge density and/or spin exchange
interactions appears to be the most natural choice.
Density correlations correspond to
an effective finite-range Coulomb interaction (repulsive or attractive)
as opposed to the on-site Hubbard $U$. On the other hand, the spin-spin
interaction is believed to be responsible for the interplay between
various spin orderings (SDW or ferromagnetism) with strong (triplet)
superconducting correlations in a variety of compounds:
Sr$_2$RuO$_4$~\cite{mackenzie_00}, UGe$_2$~\cite{saxena_00},
URhGe~\cite{aoxi_00}, ZrZn$_2$~\cite{pfeiderer_00} and
(TMTSF)$_2$X~\cite{ishiguro_00} family. Any extension of the HM breaks the
integrability, so that either approximate analytic (\eg~bosonization) or
numerical finite-system methods (\eg~Density Matrix Renormalization
Group or Quantum Monte Carlo) should be used even in 1D.
Bosonization works well in the limit of large bandwidth and predicts
several interesting effects like coexistence of spin density wave
correlations and (triplet) superconducting ones (for spin exchange
interactions~\cite{dziurzik_00,dziurzik_01}) or a transition towards
Mott-insulating phases at commensurate fillings (for charge
interactions~\cite{giamarchi}). On the other hand, in the opposite limit
of narrow-band width (atomic limit), the extended Hubbard model can be
exactly solved by means of any of the two equivalent methods - Transfer
Matrix~(TM) technique~\cite{baxter} or Composite Operator
Method~\cite{advphys,mancini_01}. In the present work we employ a $T\to 0$
extrapolation of the TM method in order to derive the zero-temperature
phase diagram of the extended HM with both nearest-neighbor~(NN) charge
and spin interactions in the narrow-band limit.
\section{Results}
The Hamiltonian of the~\nmodel~system reads as follows:
\be
   H = \sum_i \left[ -\mu n(i) +U D(i) \right]
   + V \sum_{i}n(i) n(i+1) 
   - J \sum_{i}S^z(i) S^z(i+1).
   \label{ham}
\ee
Here $n(i)\equiv n_{\upa}(i)+n_{\doa}(i)$ denotes the electron density operator
at site $i$, $D(i)\equiv n_{\upa}(i)n_{\doa}(i)$ is the double occupancy at site
$i$ while $S^{z}(i)\equiv (n_{\upa}(i)-n_{\doa}(i))/2$ denotes the $z$-component
of the spin at site $i$.
The Hamiltonian~(\ref{ham}) contains only NN term, hence the TM will be a
$4\times4$ matrix. It is easy to check that the TM elements can be
calculated through the matrix elements of an auxiliary matrix $Z$:
$T_{ij}=\exp\left( -\beta Z_{ij} \right)$, where $Z$ is defined as
follows:
\be
   Z = \left(
   \begin{array}{cccc}
	  0 & -\frac{\mu}{2} & -\frac{\mu}{2} & \frac{U}{2} -\mu \\
      -\frac{\mu}{2}  & V-\frac{J}{4} - \mu & V+\frac{J}{4} - \mu & \frac{U}{2}+2V-\frac{3}{2}\mu \\
      -\frac{\mu}{2}  & V+\frac{J}{4} - \mu & V-\frac{J}{4} - \mu & \frac{U}{2}+2V-\frac{3}{2}\mu \\
      \frac{U}{2}-\mu & \frac{U}{2}+2V-\frac{3}{2}\mu &
	  \frac{U}{2}+2V-\frac{3}{2}\mu & U+4V-2\mu
   \end{array}
   \right).
\ee
In our longer paper~\cite{long_paper} we show how to reconstruct
the $T=0$ phase diagram from the TM matrix elements. Here we apply this
method to the~\nmodel~model. We first identify all different matrix
elements (we call them energy scales). In the present case these are:
\be
   \begin{array}{ll}
	  F^0 &= 0 \\
	  F^{1/2} &= -\frac{\mu}{2}\\
	  F^{1}_1 &= \frac{U}{2}-\mu; \; F^{1}_2 = V-\frac{J}{4} - \mu; \; F^{1}_3 =V+\frac{J}{4} - \mu \\
	  F^{3/2} &= \frac{U}{2}+2V - \frac{3}{2} \mu \\
	  F^{2}   &= U+4V-2\mu.
   \end{array}
   \label{en_sc}
\ee
In~(\ref{en_sc}) we have already sorted the energy scales based on the
values of particle numbers (given by the superscript of $F$) allowed for a
given configuration. In particular, the three energy terms at $n=1$ can
be written as a unique expression as follows:
\be
   F^{1} = A-\mu,
\ee
where:
\be
   A\equiv \min\left( \frac{U}{2}, V-\frac{|J|}{4}\right) = 
   \left\{
   \begin{array}{ll}
   \frac{U}{2},& \quad \frac{U}{2}<V-\frac{|J|}{4}\\
   V-\frac{|J|}{4}, &\quad \frac{U}{2}>V-\frac{|J|}{4}.
   \end{array}
   \right.
\ee
We choose $J$ as energy scale and explore the phase diagram in the
$V-U$ plane. As $\mu$ increases, the particle number in the system
increases as well. Therefore, for each value of $n$
($n=0,\frac{1}{2},1,\frac{3}{2},2$) we can establish the ranges of the
chemical potential, within which $\mu$ can change without changing $n$.
Such ranges can be summarized as follows:
\be
    \begin{array}{llll}
	  n=0          :& \mu < \min\left( 0, A,  x, \frac{2}{3}x \right) &
	  n=1          :& \max\left( A, 2A \right) < \mu < 2x - \max\left(A,2A\right)\\
	  n=\frac{1}{2}:& 0 < \mu < \min\left( 2A,x,\frac{4}{3}x \right) & 
	  n=\frac{3}{2}:& \max\left( \frac{2}{3}x, x, 2(x-A) \right) < \mu < 2x \\
	  n=2          :& \max\left( x, \frac{4}{3} x, 2x-A, 2x \right) < \mu,
   \end{array}
   \label{ex-cond}
\ee
where we introduced $x\equiv \frac{1}{2}(U+4V)$.
Except for the cases $n=0$ and $n=2$, which are always possible, the
other cases might be incompatible with a given choice of $U,V,J$. This
occurs when the upper-bound of $\mu$ becomes lower than the lower-bound.
This condition determines the existence of the jump at a given
$n$ in the dependence $\mu(n)$. Such conditions are:
\be
    \begin{array}{ll}
	  n=\frac{1}{2} & x>0 \wedge A>0 \\ 
	  n=1           & 0 < 2A < x \textrm{ or } A<0 \wedge A<x \\
	  n=\frac{3}{2} & x>0 \wedge A>0.
   \end{array}
\ee
The conditions for the existence of jumps at $n=\frac{1}{2}$ and
$n=\frac{3}{2}$ coincide as these values of $n$ are related by the
particle-hole relation. Moreover, depending on the values of the Hamiltonian
parameters, we can distinguish four cases in the
conditions~(\ref{ex-cond}):
\vspace{0.5cm}

\begin{tabular}{llll}
   (i) $x>0$ $\wedge$ $A>0$   & (ii) $x>0$ $\wedge$ $A<0$ &
   (iii) $x<0$ $\wedge$ $A>x$ & (iv) $x<0$ $\wedge$ $A<x$.
\end{tabular}
\vspace{0.5cm}

We can identify three zones in the $V-U$ plane depending on
how many jumps there are in the dependence $\mu(n)$:
\begin{enumerate}
   \item $V>\frac{|J|}{4}$ $\wedge$ $U>0$
   \item
	  \begin{tabular}{ll}
		 (a) $\frac{U}{2}>-V-\frac{|J|}{4}$ $\wedge$ $\frac{U}{2}>V-\frac{|J|}{4}$
			$\wedge$ $V<\frac{|J|}{4}$ &
		 (b) $U<0$ $\wedge$ $V>0$ $\wedge$ $\frac{U}{2}<V-\frac{|J|}{4}$
	  \end{tabular}
   \item $V<0$ $\wedge$ $U<-2V-\frac{|J|}{2}$.
\end{enumerate}
The zone i) is characterized by the presence of jumps at all ``allowed'' values of
$n$: $n=0,\frac{1}{2},1,\frac{3}{2},2$, while in the zone ii) the jumps
are only at $n=0,1,2$. Finally, in the zone iii), the only jumps present
are at $n=0,2$. The dependence $\mu(n)$ in the three zones can be
summarized as follows:
\[
\textrm{i)}\\
\begin{array}{c|c|c|c|c}
   n=0 & n=\frac{1}{2} & n=1 & n=\frac{3}{2} & n=2 \\
   \hline
   \scriptstyle \mu <0 & \scriptstyle 0 < \mu < 2A & \scriptstyle 2A < \mu < U+4V-2A &
   \scriptstyle U+4V -2A < \mu <  U+4V & \scriptstyle U+4V < \mu
\end{array}
\]
\[
\textrm{ii)}
\begin{array}{c|c|c}
   n=0 & n=1 & n=2 \\
   \hline
   \scriptstyle \mu <A & \scriptstyle A < \mu < U+4V-A & \scriptstyle U+4V -A < \mu
\end{array}
\]
\[
\textrm{iii)}
\begin{array}{c|c}
   n=0 & n=2 \\
   \hline
   \scriptstyle \mu <\frac{U}{2}+2V & \scriptstyle \frac{U}{2}+2V < \mu
\end{array}
\]

\begin{figure*}
  \begin{tabular}{cc}
      \includegraphics[width=5cm,angle=270]{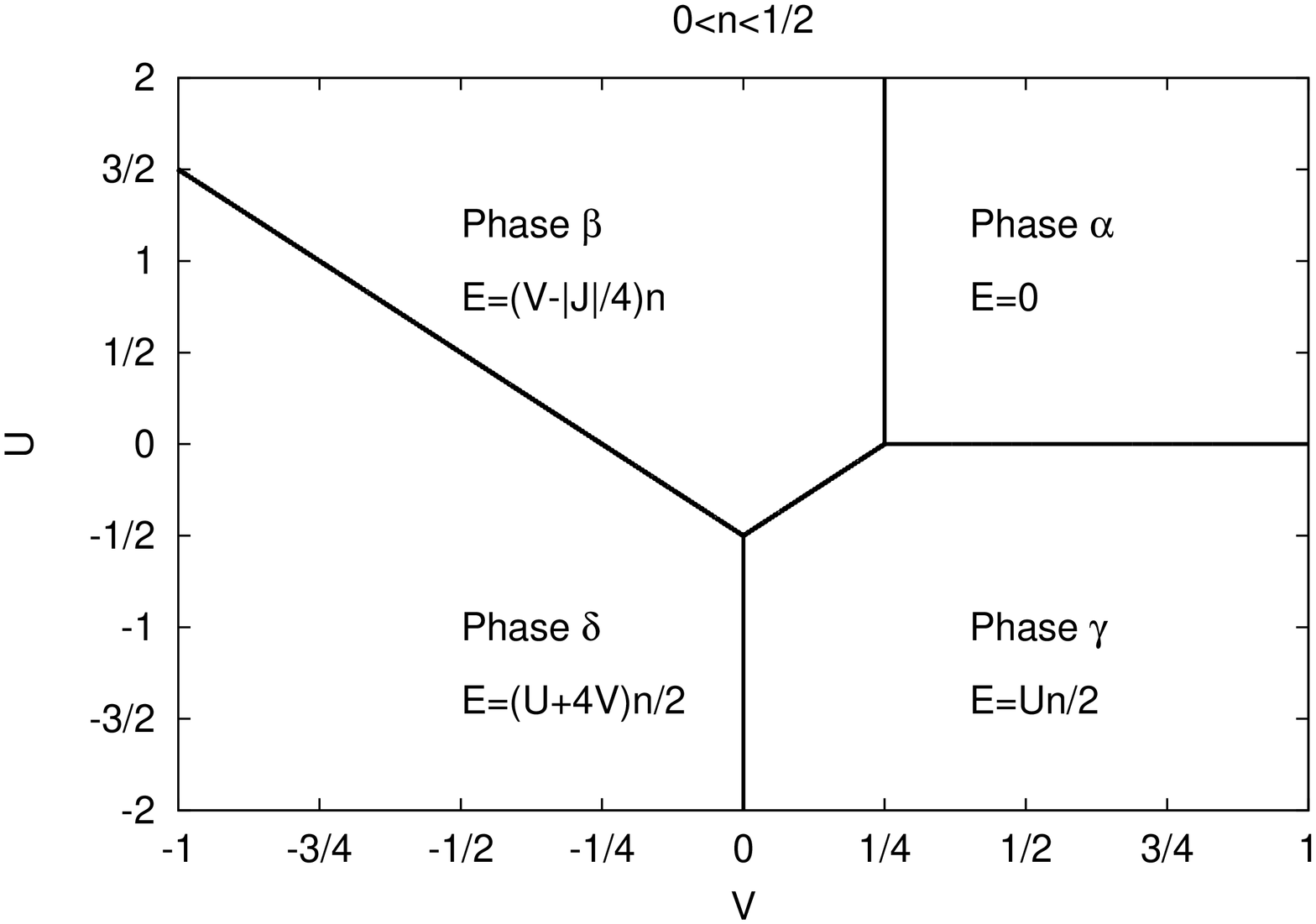}
      \includegraphics[width=5cm,angle=270]{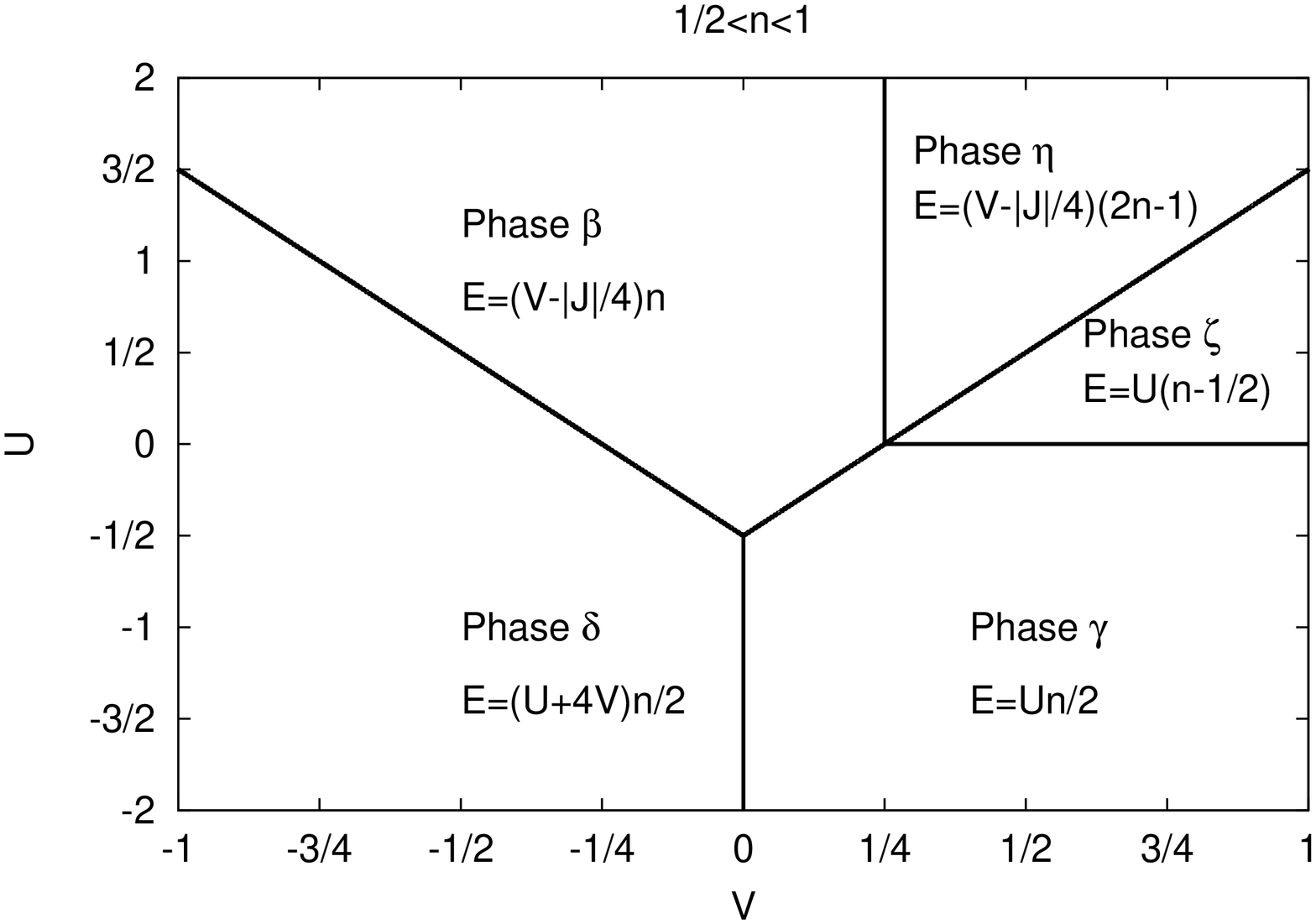}
  \end{tabular}
\caption{Left panel: Phase diagram in the case $0<n<\frac{1}{2}$.
   Right panel: Phase diagram in the case $\frac{1}{2}<n<1$.
   }
   \label{fig1}
\end{figure*}

We are now in a position to transform from canonical ($n$ fixed) to grand
canonical ($\mu$ fixed) ensemble. In the zone i) and in the range
$0<n<\frac{1}{2}$, $\mu=0$ and the free-energy per site (which is a linear
function of $n$) goes from $F_1=0$ to $F_2=-\mu/2|_{\mu=0}=0$.
Therefore, in the above interval $F$ is constantly zero and hence
the internal energy per site, which is defined as $E=\frac{1}{N}\left[\average{H}+\mu\sum_i
\average{n(i)}\right]=F+\mu n$, is zero as well.
Analogously, we can determine the behavior of the
internal energy as a function of $n$ in the whole range. These findings
for the zones i) and ii) are summarized in Tables~\ref{tab1}
and~\ref{tab2}. In the zone iii), it is easy to verify that
$E=(U+4V)\frac{n}{2}$ for all $0<n<2$.

\begin{table}
   \caption{Internal energy as a function of particle concentration $E(n)$ for
   the Zone i).}
   \centerline{
   \begin{tabular}{c|c|c|c}
	  \hline
	  $0<n<\frac{1}{2}$ & $\frac{1}{2} <n<1$ & $1<n<\frac{3}{2}$ & $\frac{3}{2}<n<2$ \\
	  \hline
	  $E=0$ & $E=A(2n-1)$ & $E=(U+4V)(n-1) + A(3-2n)$ & $E=(U+4V)(n-1)$\\
	  \hline
   \end{tabular}
   }
   \label{tab1}
\end{table}
\begin{table}
   \caption{Internal energy as a function of particle concentration $E(n)$ for
   the Zone ii).}
   \centerline{
   \begin{tabular}{c|c}
	  \hline
	  $0<n<1$ & $1<n<2$ \\
	  \hline
	  $E=An$ & $E=(U+4V)(n-1)+A(2-n)$ \\
	  \hline
   \end{tabular}
   }
   \label{tab2}
\end{table}

By definition, $A$ takes two different forms depending on whether $\frac{U}{2}$ is
greater or not than $V-\frac{|J|}{4}$. This means that everywhere
$A$ appears, the two cases should be considered and, therefore, the zones
i) and ii) will be further divided in two by the line
$\frac{U}{2}=V-\frac{|J|}{4}$. Thanks to the particle-hole symmetry enjoyed by
the system, we can consider the phase diagrams just for $n\leqslant 1$.
Two cases can be distinguished: $0<n<\frac{1}{2}$ and $\frac{1}{2}<n<1$,
together with the particular cases $n=\frac{1}{2}$ and $n=1$. The
two phase diagrams are depicted in Fig.~\ref{fig1}. We proceed now to
the description of the different phases appearing in the whole phase
diagram.

Phase~$\alpha$ is located in the range $V>\frac{|J|}{4}\wedge U>0$ and
$0<n<\frac{1}{2}$. The internal energy is zero because the particles are
are far apart to activate any of the Hamiltonian terms. The chemical
potential vanishes, $\mu=0$. When $n>\frac{1}{2}$, the former Phase $\alpha$
gives rise to two new phases ($\eta$ and $\zeta$).

Phase~$\beta$ is located in the range $\frac{U}{2}>|V|-\frac{|J|}{4} \wedge V<\frac{|J|}{4}$.
Energy: $E=(V-\frac{|J|}{4})n$ and thus in this phase the singly
occupied sites at NN distance interact via density-density and spin-spin
interactions. The spin-spin correlation function takes the value:
\be
   \average{S^z(i)S^z(i+1)} = -\frac{\partial E}{\partial J} = 
   \frac{\sign(J)n}{4}.
\ee
Hence, if $J<0$, the system is dominated by antiferromagnetic
correlations, while if $J>0$ the correlations are ferromagnetic.
This phase exists in the range $0<n<1$. Chemical potential $\mu=V-\frac{|J|}{4}$,
NN charge-density correlations $\average{n(i)n(i+1)}=n$. This phase
exhibits spontaneous magnetization, uniform in case $J>0$, or staggered in case $J<0$.

Phase~$\gamma$ is located in the range $V>0\wedge U<0\wedge
\frac{U}{2}<V-\frac{|J|}{4}$. Energy is $E=\frac{U}{2}n$, and therefore
in this phase the doubly occupied sites are separated by at least one
empty site. This phase exists in the range $0<n<1$. Chemical potential $\mu=\frac{U}{2}$.
The only non-zero correlation function is $\average{D(i)}=n/2$.

Phase~$\delta$ is located in the range $V<0\wedge \frac{U}{2}<-V-\frac{|J|}{4}$
for all $0<n<2$. The energy: $E=(U+4V)\frac{n}{2}$ and thus, in this phase, the
doubly occupied sites are placed at the NN distance. Chemical potential
$\mu=\frac{U}{2}+2V$.
The only non-zero correlation functions are $\average{D(i)}=n/2$,
$\average{n(i)n(i+1)}=2n$ and $\average{D(i)D(i+1)}=n/2$.

Phase~$\eta$ is located in the range $V>\frac{|J|}{4} \wedge \frac{U}{2}>V-\frac{|J|}{4}$
and $\frac{1}{2}<n<1$. The energy $E=(V-\frac{|J|}{4})(2n-1)$. This is
the second phase with magnetic correlations and is similar to the
adjacent Phase $\beta$ in the sense that in both phases density-density
and spin-spin correlations are present, although the expressions for the
correlation functions are different: $\average{n(i)n(i+1)}=2n-1$,
$\average{S^z(i)S^z(i+1)}=\frac{\sign(J)}{4}(2n-1)$. Chemical potential
$\mu=2(V-\frac{|J|}{4})$. There also exists spontaneous magnetization,
uniform in case $J>0$, or staggered in case $J<0$.

Phase~$\zeta$ is located in the range $\frac{U}{2}>V-\frac{|J|}{4} \wedge U>0$
and $\frac{1}{2}<n<1$. The energy $E=U(n-\frac{1}{2})$. This phase is similar to
the Phase $\gamma$ in the sense explained above. Chemical potential
$\mu=U$, while the only non-vanishing correlation function is
$\average{D(i)}=n-1/2$.

The phase boundaries for the phases at $n>1$ are identical to those at
$n<1$ and the expressions for the internal energy can be obtained
from those at $n<1$ by invoking the particle-hole symmetry:
\be
   E(2-n) = E(n) + (U+4V)(1-n), 
\ee
while for the chemical potential the particle-hole relation states:
\be
   \mu(2-n) = (U+4V)-\mu(n).
\ee
\section{Conclusions}
In this work we present an original method for extrapolating
the TM technique at $T=0$ and use it to obtain the exact $T=0$
phase diagram of the extended Hubbard~\nmodel~model in the narrow-band
limit. Depending on the values of the Hamiltonian parameters, the
orderings involve double occupancy, NN density and spin correlations. We
note that the phase diagram boundaries do not depend on the sign of $J$,
although the spin correlations do. The spin correlations appear to be
rather fragile in our model: the Ising term in the Hamiltonian acts only
if the NN sites are singly occupied. However, in the same conditions the
$V$ term is also active, while the $U$ term tends to create either
doubly occupied or empty sites (depending on the sign of $U$). That is
why, when either $V$ is large and positive or $U$ is large and negative
the spin correlations are completely suppressed. The introduction of $J$
affects only the phases ($\beta$ and $\eta$) where the sites are singly
occupied. In these phases, for $J=0$~\cite{mancini_00, mancini_01},
the spins are not ordered while, if $J\ne 0$, a ferromagnetic or
antiferromagnetic order is established depending on the sign of $J$.
\section{Acknowledgements}
We acknowledge the CINECA award under the ISCRA initiative (project
BSMOSCES), for the availability of high performance computing resources
and support.

\section*{References}

\providecommand{\newblock}{}

\end{document}